\documentclass[secnumarabic,prl,10pt,notitlepage,footinbib,nobibnotes,showpacs,onecolumn]{revtex4}
\usepackage{amsmath}
\usepackage{graphicx}

\input{tcilatex}

\begin{document}

\title{Realization of decoherence-free subspace using Multiple-Quantum
coherences}
\author{Daxiu Wei}
\email{dxwei@wipm.ac.cn}
\affiliation{State Key Laboratory of Magnetic Resonance and Atomic and Molecular Physics,
Wuhan Institute of Physics and Mathematics, Chinese Academy of Sciences,
Wuhan 430071, P. R. China}
\author{Jun Luo, Xianping Sun, Xizhi Zeng, Mingsheng Zhan}
\affiliation{State Key Laboratory of Magnetic Resonance and Atomic and Molecular Physics,
Wuhan Institute of Physics and Mathematics, Chinese Academy of Sciences,
Wuhan 430071, P. R. China}
\author{Maili Liu }
\email{ml.liu@wipm.ac.cn}
\affiliation{State Key Laboratory of Magnetic Resonance and Atomic and Molecular Physics,
Wuhan Institute of Physics and Mathematics, Chinese Academy of Sciences,
Wuhan 430071, P. R. China}

\begin{abstract}
This letter presents a two-dimensional nuclear magnetic resonance(NMR)
approach for constructing a two-logical-qubit decoherence-free subspace
(DFS) based on the fact that the three protons in a CH$_{3}$ spin system can
not be resolved in one-dimension NMR spectroscopy, but to a certain extent,
can be distinguished by two-dimensional multiple-quantum NMR. We used four
noisy physical nuclear spins, including three protons and one carbon in the
CH$_{3}$ spin system, to generate two decoherence-free logical qubits. It
made full use of the unaddressed spins which could not be used in
one-dimensional spectrum. Furthermore, we have experimentally demonstrated
such an approach. Our experimental results have shown that our DFS can
protect against far more types of decoherence than the one composed of four
noisy physical qubits all with different chemical shifts. More importantly,
this idea may provide new insights into extending qubit systems in the sense
that it effectively utilizes the magnetically equivalent nuclei.
\end{abstract}

\maketitle

Quantum computers could, in theory, be superior to classical computers when
performing a number of computational tasks, such as factorizing a large
number \cite{shor}, searching unsorted database \cite{grover} and,
especially, simulating quantum systems themselves \cite{lloyd}. Among
various techniques, liquid nuclear magnetic resonance (NMR) approach is,
currently, the most realistic tool \cite{jones}. If the chemical shift of a
nuclear spin is well resolved and can be controlled by selective
radio-frequency (RF) pulses without disturbing the other spins, the spin can
be used as a qubit for quantum computing. Although NMR quantum computing has
made the greatest progress towards QIP \cite{jones}, it is very difficult to
scale up current NMR quantum computers to large sizes (more than 10 qubits)%
\cite{warren}. There are many obstacles that limit wide applications of NMR
in quantum computation. Firstly, in addition to the low signal-to-noise
ratio (SNR) in NMR experiments, the SNR decreases exponentially with
increasing of qubit numbers. Secondly, there are only few types of nuclear
spins which might be used as qubits, such as $^{1}$H, $^{13}$C, $^{15}$N, $%
^{19}$F, and $^{31}$P etc. Thirdly, because of limitations in the number of
the receive channels and the transmit channels in the commercial NMR
spectrometers, one always had to employ homonuclear systems and selectively
control each of the spins when implementing multi-qubit quantum computing.
Finally, frequency dispersion limits the number of usable homonuclear spins
or qubits. In the case of proton NMR at 500 MHz, the chemical shift range is
generally less than 5000 Hz. This small spectral window restricts the number
of qubits in a homonuclear system. In addition, it is extremely hard to
select or design a special molecule whose protons all have well resolved
chemical shifts and are coupled to each other. It is thus a great challenge
for NMR spectroscopists to utilize all of the available spins in the real
molecule. It has been noticed that magnetically equivalent spins, such as
the three protons in a methyl, are used as a single physical qubit. One can
wonder whether it is possible to use the equivalent spins as individual
qubits. In this context, we will describe a two-dimensional (2D) NMR
technique to implement logical qubit by making full use of the three protons
in the methyl.

The new concept of using 2D NMR to perform quantum computing was first
proposed by M\'{a}di and Ernst \textit{et al}. in 1998\cite{madi}. The
quantum logical gates and Deutsch-Jozsa quantum algorithm were also
experimentally realized in 2D NMR with the help of spin- and transition-
selective pulses\cite{kumar1,kumar2}. In these experiments, since single
quantum transitions were used in both dimensions, the equivalent protons are
indistinguishable and can only be utilized as one physical qubit. On the
contrary, using 2D multiple-quantum NMR J-resolved spectroscopy (MQ-JRES) of
the spins in alanine\cite{liu}, we observed that the methyl has four
resolved multi-quantum (MQ) coherences which have no interchange during the
free evolution time. The 2D MQ-JRES spectrum can be explained in detail by a
SI$_{3}$-M spin system (for the $^{13}$CH$_{3}$-$^{12}$CH in alanine). After
four spins are excited to their highest quantum state, the SI$_{3}$ spin
system will include four coherences (one quadruple quantum, two double
quanta, and one zero quantum)\cite{liu}. Moreover, different multi-quantum
coherences are modulated differently by their coupling constants to the
remote spin M, and is distinguished by projecting the 2D multi-quantum onto
the F1 dimension. In this case, the three protons in the methyl play the
roles of three physical qubits. The following question arises:
\textquotedblleft can three physical qubits composed of three protons in a
methyl be used to implement the 2D NMR quantum
computing(QC)?\textquotedblright\ This topic will be discussed and solved in
the following context.

Here, we will describe how to use three protons and a carbon in a methyl as
four physical qubits to produce four multiple quantum coherences, and then
utilize these multiple quantum coherences as logical qubits to construct an
operator subspace DFS in a Liouville space\cite{madi,bru,jones} by virtue of
the 2D multi-quantum J-resolved NMR spectrum (MQ-JRES). We also present
experimental results in demonstrating this novel idea.

There are two kinds of passive error control codes for counteracting the
decoherence. One is decoherence-free subspace (DFS) \cite{pzan,duan,lidar}
and the other is noiseless subsystems (NSs)\cite{vio,sde,Knill-NS,pz}. They
had been widely studied in both theoretically and experimentally \cite%
{pzan,duan,lidar,vio,sde,Knill-NS,dfs,lor}. Recently, the two-logic-qubit
DFS composed of four physical qubits were demonstrated in NMR quantum
computing\cite{dfs}. All states in DFS are immune to the error operation E$%
_{d}$ with the form $E_{d}=\alpha _{d,0}E_{1}E_{2}E_{3}E_{4}+\alpha
_{d,1}X_{1}X_{2}E_{3}E_{4}+\alpha _{d,2}E_{1}E_{2}X_{3}X_{4}+\alpha
_{d,3}X_{1}X_{2}X_{3}X_{4}$, where \textit{E} is the unit operator and 
\textit{X}$_{i}$ (i=1, 2, 3, and 4) are operators which make the physical
qubits flipped. Just as the other methods to avoid the decoherence problem,
the construction of the DFS always needs at least two physical qubits to
produce one logical qubit.

In order to express a DFS as state vectors or density matrices, it is
instructive to discuss the connection between logical-qubit state vectors
and density matrices.\ For simplicity, we first consider a DFS with two
physical qubits\cite{won}, which protects against the error set \{$%
E_{1}E_{2} $, $X_{1}X_{2}$, $Z_{1}Z_{2}$\} in the form of

\begin{eqnarray}
|0\rangle _{L}^{0} &\text{=}&\frac{1}{\sqrt{2}}(|01\rangle +|10\rangle , 
\notag \\
|1\rangle _{L}^{0} &\text{=}&\frac{1}{\sqrt{2}}(|01\rangle -|10\rangle \text{%
.}
\end{eqnarray}%
The above states can also be rewritten in the form of density operators and
the state $|0\rangle _{L}^{0}$ corresponds to the following expression, 
\begin{eqnarray}
\rho _{(L0)}^{L} &=&|0\rangle _{LL}^{00}\langle 0|  \notag \\
&=&\frac{1}{2}(\frac{1}{2}%
-2I_{z}^{1}I_{z}^{2}+2I_{x}^{1}I_{x}^{2}+2I_{y}^{1}I_{y}^{2})\text{.}
\end{eqnarray}%
On the other hand, an experimental density matrix ($\rho $) could be
separated into different orders \textit{p }( such as zero-, single-, two-,
or quadruple-quantum coherence and so on)\cite{ernst},%
\begin{equation*}
\rho =\dsum\limits_{p=-2l}^{+2l}\rho ^{p},
\end{equation*}%
where $l$ is the maximum total spin quantum number. For example, we can
decompose Eq. 2 into the superposition of zero quantum and double quantum, 
\begin{equation}
\rho _{(L0)}^{L}=\rho ^{0}+\rho ^{2}+\rho ^{-2}=\underset{\rho ^{0}}{%
\underbrace{\frac{1}{4}-I_{z}^{1}I_{z}^{2}}}+\underset{\rho ^{2}+\rho ^{-2}}{%
\underbrace{I_{x}^{1}I_{x}^{2}+I_{y}^{1}I_{y}^{2}}}\text{.}
\end{equation}%
where $I_{z}$, $I_{x}$, $I_{y}$ are product operators with $I_{z}$=$\frac{1}{%
2}\sigma _{z}$=$\frac{1}{2}\left[ 
\begin{array}{cc}
1 & 0 \\ 
0 & -1%
\end{array}%
\right] $, $I_{x}$=$\frac{1}{2}\sigma _{x}$=$\frac{1}{2}\left[ 
\begin{array}{cc}
0 & 1 \\ 
1 & 0%
\end{array}%
\right] $, $I_{y}$=$\frac{1}{2}\sigma _{y}$=$\frac{i}{2}\left[ 
\begin{array}{cc}
0 & -1 \\ 
1 & 0%
\end{array}%
\right] $. In a NMR experiment the different orders will evolve according to
their own rules. Actually, the different orders$\ $are the elementary
constituents of the density matrices. In this case, it is possible to create
a DFS just by using the different orders as basis states in the density
operator representation. Therefore, we should first find a group of logical
orthonormal basis states which span a self-contained space. Such a group of
basis states may be expressed in the form of density operators just as the
four logical states proposed in Ref.\cite{dfs}. But, on the other hand, it
could also exist in an operator subspace and be presented in the fashion of
density operators as long as one can find a group of logical orthonormal
basis states.

In looking for such an operator subspace DFS, we noted that there existed
four MQ coherences in the spin system SI$_{\text{3}}$ (such as the methyl $%
^{13}$CH$_{3}$ where S denotes $^{13}$C and I stands for $^{1}$H), including
one quadruple quantum (QQ), two double quantum (DQ$1,$DQ$2$) and one zero
quantum (ZQ). More importantly, these four MQ coherences can be
distinguished by applying Liu \textit{et al}.'s pulse sequence\cite{liu}.
Their product operator forms are

\begin{eqnarray}
\rho _{1}(\text{QQ})
&=&3I_{x}I_{y}I_{y}S_{y}-3I_{x}I_{x}I_{y}S_{x}-I_{x}I_{x}I_{x}S_{y}+I_{y}I_{y}I_{y}S_{x},
\notag \\
\rho _{2}(\text{DQ1})
&=&3I_{x}I_{y}I_{y}S_{y}+3I_{x}I_{x}I_{y}S_{x}-I_{x}I_{x}I_{x}S_{y}-I_{y}I_{y}I_{y}S_{x},
\notag \\
\rho _{3}(\text{DQ2})
&=&I_{x}I_{y}I_{y}S_{y}+I_{x}I_{x}I_{y}S_{x}+I_{x}I_{x}I_{x}S_{y}+I_{y}I_{y}I_{y}S_{x},
\notag \\
\rho _{4}(\text{ZQ})
&=&I_{x}I_{y}I_{y}S_{y}-I_{x}I_{x}I_{y}S_{x}+I_{x}I_{x}I_{x}S_{y}-I_{y}I_{y}I_{y}S_{x}.
\end{eqnarray}%
It is obvious that not all of the above four MQ coherences are orthogonal to
each other, namely, they could not be used as orthonormal basis states for a
four-dimension operator space. Nevertheless, if we slightly modify them to
the following set of operators shown in Eq. 5, it is interesting to find
that the four modified density operators (they also belong to multiple
quantum coherences) span a two-logical-qubit DFS. We denote the four
modified MQ coherences as\textbf{\ }$\rho _{1}$, $\rho _{2}$, $\rho _{3}$, $%
\rho _{4}$, which correspond to the logical states $|00\rangle _{LL}\langle
00|$, $|01\rangle _{LL}\langle 01|$, $|10\rangle _{LL}\langle 10|$, $%
|11\rangle _{LL}\langle 11|$, respectively. These states are written as

\begin{eqnarray}
\rho _{1} &=&|00\rangle _{LL}\langle
00|=I_{x}I_{y}I_{y}S_{y}-I_{x}I_{x}I_{y}S_{x}-I_{x}I_{x}I_{x}S_{y}+I_{y}I_{y}I_{y}S_{x},
\notag \\
\rho _{2} &=&|01\rangle _{LL}\langle
01|=I_{x}I_{y}I_{y}S_{y}+I_{x}I_{x}I_{y}S_{x}-I_{x}I_{x}I_{x}S_{y}-I_{y}I_{y}I_{y}S_{x},
\notag \\
\rho _{3} &=&|10\rangle _{LL}\langle
10|=I_{x}I_{y}I_{y}S_{y}+I_{x}I_{x}I_{y}S_{x}+I_{x}I_{x}I_{x}S_{y}+I_{y}I_{y}I_{y}S_{x},
\notag \\
\rho _{4} &=&|11\rangle _{LL}\langle
11|=I_{x}I_{y}I_{y}S_{y}-I_{x}I_{x}I_{y}S_{x}+I_{x}I_{x}I_{x}S_{y}-I_{y}I_{y}I_{y}S_{x},
\end{eqnarray}%
where the normalization constants have been omitted. Specifically, the
states corresponding to these density operators are the eigenstates of the
operator $E_{n}$,

\begin{eqnarray}
E_{n} &\text{=}&\alpha _{n,0}E^{1}E^{2}E^{3}E^{4}\text{+}\alpha
_{n,1}E^{1}E^{2}E^{3}Z^{4}\text{+}\alpha _{n,2}Z^{1}Z^{2}Z^{3}E^{4}\text{+}%
\alpha _{n,3}Z^{1}Z^{2}Z^{3}Z^{4}\text{+}  \notag \\
&&\alpha _{n,4}X^{1}X^{2}X^{3}X^{4}\text{+}\alpha _{n,5}X^{1}X^{2}X^{3}Y^{4}%
\text{+}\alpha _{n,6}Y^{1}Y^{2}Y^{3}X^{4}\text{+}\alpha
_{n,7}Y^{1}Y^{2}Y^{3}Y^{4}\text{.}
\end{eqnarray}%
In a word, by using the spin system SI$_{\text{3}}$ we have actually
constructed a two-logical-qubit DFS which can resist the error $E_{n}$. Now
the next work is to find an effective experimental technique of realizing
the DFS with the above product operators.

Our method to implement the above DFS is based on the idea in Ref.\cite{liu}%
, in which the pulse sequence can be divided into three parts. The first
part (the pulse sequences before the encoding gradient pulse $G_{e}$) is to
generate the highest state $8I_{x}I_{y}I_{y}S_{y}$ which consists of the
above four coherences. In the second part (including the gradient pulse $%
G_{e}$, the evolution time t$_{1}$ and the two 180$^{\circ }$ pulses on
spins \textit{I} and \textit{S}), each of the above four coherences evolves
differently. At the end of the second part, the four coherences have the
following forms

\begin{eqnarray}
\rho 
%TCIMACRO{\U{b4}}%
%BeginExpansion
{\acute{}}%
%EndExpansion
_{1}(\text{QQ})
&=&(3I_{x}I_{y}I_{y}S_{y}-3I_{x}I_{x}I_{y}S_{x}-I_{x}I_{x}I_{x}S_{y}+I_{y}I_{y}I_{y}S_{x})
\notag \\
&&\cos [\pi (3J_{IM}+J_{SM})t_{1}]\exp (-t_{1}/T_{2QQ}),  \notag \\
\rho 
%TCIMACRO{\U{b4}}%
%BeginExpansion
{\acute{}}%
%EndExpansion
_{2}(\text{DQ1})
&=&(3I_{x}I_{y}I_{y}S_{y}+3I_{x}I_{x}I_{y}S_{x}-I_{x}I_{x}I_{x}S_{y}-I_{y}I_{y}I_{y}S_{x})
\notag \\
&&\cos [\pi (3J_{IM}-J_{SM})t_{1}]\exp (-t_{1}/T_{2DQ1}),  \notag \\
\rho 
%TCIMACRO{\U{b4}}%
%BeginExpansion
{\acute{}}%
%EndExpansion
_{3}(\text{DQ2})
&=&(I_{x}I_{y}I_{y}S_{y}+I_{x}I_{x}I_{y}S_{x}+I_{x}I_{x}I_{x}S_{y}+I_{y}I_{y}I_{y}S_{x})
\notag \\
&&\cos [\pi (J_{IM}+J_{SM})t_{1}]\exp (-t_{1}/T_{2DQ2}),  \notag \\
\rho 
%TCIMACRO{\U{b4}}%
%BeginExpansion
{\acute{}}%
%EndExpansion
_{4}(\text{ZQ})
&=&(I_{x}I_{y}I_{y}S_{y}-I_{x}I_{x}I_{y}S_{x}+I_{x}I_{x}I_{x}S_{y}-I_{y}I_{y}I_{y}S_{x})
\notag \\
&&\cos [\pi (J_{IM}-J_{SM})t_{1}]\exp (-t_{1}/T_{2ZQ}).
\end{eqnarray}%
Therefore, the four MQ coherences can be distinguished by the remote J
couplings of the spins I and S with the spin M, respectively, and the
transverse relaxations. The third part is to transform the multiple-quantum
coherences into the detectable signals. It should be noted that different
multiple-quantum coherences have different projection onto the F$_{1}$
dimension, which means that the resonances of the different multiple-quantum
are resolved. In particular, the selectively detected MQ-JRES spectrum can
be achieved by adjusting the strength ratio between the encoding and
decoding gradient pulses(i.e. between $G_{e}$ and $G_{d}$). One example is
that when the gradient strength ratio $\frac{G_{e}}{G_{d}}$ is -8:10, the
double quantum coherence$($DQ2$)$ could be selectively detected. It is also
worthwhile noting that there is no interchange between the multi-quantum
coherences during the evolution period.

Considering that product operators $\rho _{i}$ in Eq. $(4)$ and $\rho _{i}($%
MQ$)$ in Eq. $(5)$ have the similar forms, the states $\rho _{1},\rho
_{2},\rho _{3},\rho _{4}$ could be obtained by using special pulse sequences
similar to the one in Ref.\cite{liu}. That is, the method used to yield the
states $\rho _{3}$ and $\rho _{4}$ is, respectively, the same as the one
used to produce the multi-quantum coherences $\rho _{3}($DQ2$)$ and $\rho
_{4}($ZQ$)$. The states $\rho _{1}$ and $\rho _{2}$ can also be created by
applying the pulse sequences corresponding to the multi-quantum coherences $%
\rho _{1}($QQ$)$ and $\rho _{2}($DQ1$)$ apart from adding two special-angle
pulses on spins C and H, respectively.

Interestingly, due to the same chemical shift of the three protons in a
methyl, error operators only affect them simultaneously, and the following
error operator E$_{m}$ doesn't exist:

\begin{eqnarray}
E_{m} &=&\alpha _{m,0}\sigma ^{1}E^{2}E^{3}E^{4}\text{+}\alpha
_{m,1}E^{1}\sigma ^{2}E^{3}E^{4}\text{+}\alpha _{m,2}E^{1}E^{2}\sigma
^{3}E^{4}\text{+}\alpha _{m,3}\sigma ^{1}\sigma 
%TCIMACRO{\U{b4}}%
%BeginExpansion
{\acute{}}%
%EndExpansion
^{2}E^{3}E^{4}\text{+}  \notag \\
&&\alpha _{m,4}\sigma ^{1}E^{2}\sigma 
%TCIMACRO{\U{b4}}%
%BeginExpansion
{\acute{}}%
%EndExpansion
^{3}E^{4}\text{+}\alpha _{m,5}E^{1}\sigma ^{2}\sigma 
%TCIMACRO{\U{b4}}%
%BeginExpansion
{\acute{}}%
%EndExpansion
^{3}E^{4}\text{+}\alpha _{m,6}\sigma ^{1}E^{2}E^{3}\sigma 
%TCIMACRO{\U{b4}}%
%BeginExpansion
{\acute{}}%
%EndExpansion
^{4}\text{+}\alpha _{m,7}E^{1}E^{2}\sigma ^{3}\sigma 
%TCIMACRO{\U{b4}}%
%BeginExpansion
{\acute{}}%
%EndExpansion
^{4}  \notag \\
&&\text{+}\alpha _{m,8}E^{1}\sigma ^{2}E^{3}\sigma 
%TCIMACRO{\U{b4}}%
%BeginExpansion
{\acute{}}%
%EndExpansion
^{4}\text{+}\alpha _{m,9}\sigma ^{1}\sigma 
%TCIMACRO{\U{b4}}%
%BeginExpansion
{\acute{}}%
%EndExpansion
^{2}E^{3}\sigma 
%TCIMACRO{\U{b4}}%
%BeginExpansion
{\acute{}}%
%EndExpansion
%TCIMACRO{\U{b4}}%
%BeginExpansion
{\acute{}}%
%EndExpansion
^{4}\text{+}\alpha _{m,10}\sigma ^{1}E^{2}\sigma 
%TCIMACRO{\U{b4}}%
%BeginExpansion
{\acute{}}%
%EndExpansion
^{3}\sigma 
%TCIMACRO{\U{b4}}%
%BeginExpansion
{\acute{}}%
%EndExpansion
%TCIMACRO{\U{b4}}%
%BeginExpansion
{\acute{}}%
%EndExpansion
^{4}\text{+}\alpha _{m,11}E^{1}\sigma ^{2}\sigma 
%TCIMACRO{\U{b4}}%
%BeginExpansion
{\acute{}}%
%EndExpansion
^{3}\sigma 
%TCIMACRO{\U{b4}}%
%BeginExpansion
{\acute{}}%
%EndExpansion
%TCIMACRO{\U{b4}}%
%BeginExpansion
{\acute{}}%
%EndExpansion
^{4},
\end{eqnarray}%
where $\sigma $, $\sigma 
%TCIMACRO{\U{b4}}%
%BeginExpansion
{\acute{}}%
%EndExpansion
$, $\sigma 
%TCIMACRO{\U{b4}}%
%BeginExpansion
{\acute{}}%
%EndExpansion
%TCIMACRO{\U{b4}}%
%BeginExpansion
{\acute{}}%
%EndExpansion
\in \{X,Y,Z\}$. In other words, the DFS we constructed can avoid errors E$%
_{n}$ and E$_{m}$ simultaneously. However, that one proposed by Ollerenshaw 
\textit{et al.} based on four different physical qubits only protests
against small parts of errors $E_{n}$ and $E_{m}$. It implies that it is
more effective to adopt the SI$_{\text{n}}$ system to realize the DFS. On
the other hand, in 1D NMR quantum computing, the spins I$_{\text{n}}$ in SI$%
_{\text{n}}$ (n=2, 3) were always regarded as one qubit, so in a sense we
actually implemented a DFS in a two-qubit physical quantum system, which has
more advantages over the one constructed by four physical qubits.

The experiment was carried out on a Varian INOVA 600 spectrometer. The
sample was 20mg alanine in 0.5ml D$_{2}$O, with one carbon in methyl
labelled. The molecular structure is $^{13}$CH$_{3}$-$^{12}$CH(NH$_{2}$)-$%
^{12}$COOH. The spin system SI$_{\text{3}}$M presents $^{13}$CH$_{3}$-$^{12}$%
CH in the alanine molecule. The coupling constants obtained from the
high-resolution 1D NMR spectra, $J_{\text{SI}}$, $J_{\text{SM}}$ and $J_{%
\text{IM}}$ are 129.8Hz, 4.5Hz and 7.3Hz, respectively. Using pulsed field
gradients, which is similar to the method in Ref.\cite{liu}, we can
selectively obtain the states $\rho _{1},\rho _{2},\rho _{3}$ or $\rho _{4}$%
. For example, $\rho _{3}$ can be selectively detected by applying the pulse
sequences shown in Fig.1. Here, the required gradient strength ratio between
the encoding gradient($G_{e}$) and the decoding gradient ($G_{d}$) is -8:10.
The delay $\bigtriangleup $ is 3.88ms corresponding to the coupling
constants $J_{\text{SI}}$=129.8Hz. The spectral widths are 30 and 4000 Hz in
the F$_{1}$ and F$_{2}$ dimensions, respectively. The data points in the F$%
_{1}$ and F$_{2}$ dimensions are 64 and 10000. In order to show that this
DFS' protection against the errors $E_{n}$ and $E_{m}$, we performed a
series of errors belonging to $E_{n}$ and $E_{m}$ at the position c\ in
Fig.1. Whether the DFS constructed by two-dimensional MQ J-resolved spectrum
protects against errors $E_{n}$ and $E_{m}$ or not, is demonstrated by
observing and comparing the 2D NMR spectra before and after applying the
decoherence operators.

Typical experimental results are shown in Fig.2, where (A), (B) correspond
to the spectra of $\rho _{3}$ before and after one performs the error
operator X$^{1}$X$^{2}$X$^{3}$Y$^{4}$, respectively. It is interesting to
note that the spectrum in (A) is the same as the one in (B). This means that 
$\rho _{3}$ is certainly not affected by this decoherence. Similar outcomes
are\textit{\ }also obtained when the other operators belonging to $E_{n}$\
or $E_{m}$ are applied to $\rho _{1},\rho _{2},\rho _{3}$ or $\rho _{4}$.
Due to limitation in length, we do not reproduce the experimental results.

Decoherence has been one of the severe impediments in quantum information
processing. Many techniques have been widely studied. Therefore it is
significative to experimentally and theoretically focus on the DFS that has
been developed as a tool for protecting QIP. The idea for constructing the
DFS provided in this paper shows new insights into effectively utilizing the
spin system including CH$_{\text{n}}$ (n=2,3) to achieve quantum computing
as well as extending the number of qubits. This process is impossible in 1D
NMR, since the magnetically equivalent spins in CH$_{\text{n}}$ can not been
fully used. But here we showed that 2D MQ-JRES could do it by using the MQ
coherences which were modulated by the remote spin. In addition, a novel
point is that since the n numbers of H have the same chemical shift, the
errors only affect them together, so to this extent\textit{, }this DFS may
avoid more errors than others that were constructed by spins with different
chemical shifts.

Simultaneously, when one performs 1D NMR quantum computing with a sample
containing the CH$_{\text{n}}$ group, all the protons in each CH$_{\text{n}}$
group which could only be used as one qubit, couple to the other spins. In
this condition, the spectra of the other spins become more complex than
those for which there are no magnetically equivalent spins in the sample,
although their number of qubits is the same. So it seems that the CH$_{\text{%
n}}$ group just makes 1D NMR QC more complex. However, just as described
above, the CH$_{\text{n}}$ group, in 2D MQJES, makes our DFS more powerful.
Here, we merely performed a special DFS applying 2D MQJES. The demonstration
of some quantum algorithms based on this DFS is under way.

In summary, we have provided a new two-logical-qubit DFS with 2D MQ-JRES and
the three magnetically equivalent spins in CH$_{3}$. In a sense, the three
protons are distinguishable. The corresponding experimental demonstration
has been completed. Compared to the other schemes for implementing DFS, our
method protected against more error operators just by employing a smaller
nuclear system. Specifically, this idea can also be applied to implement
other QIP tasks on NMR QC. It follows that when the selected spin system
contains a methyl, we can implement an (n+2)-qubit quantum computing just
using n qubits of one-dimensional NMR QC. Conclusively, the 2D NMR method
presented in this letter may be helpful for developing the multi-qubit NMR
QC.

This work has been supported by the National Natural Science Foundation of
China under Grant No. 10374103, the National Basic Research Programme of
China under Grant No.2001CB309309, and the National Knowledge Innovation
Program of the Chinese Academy of Sciences under Grant No: KJCX2-W1-3.

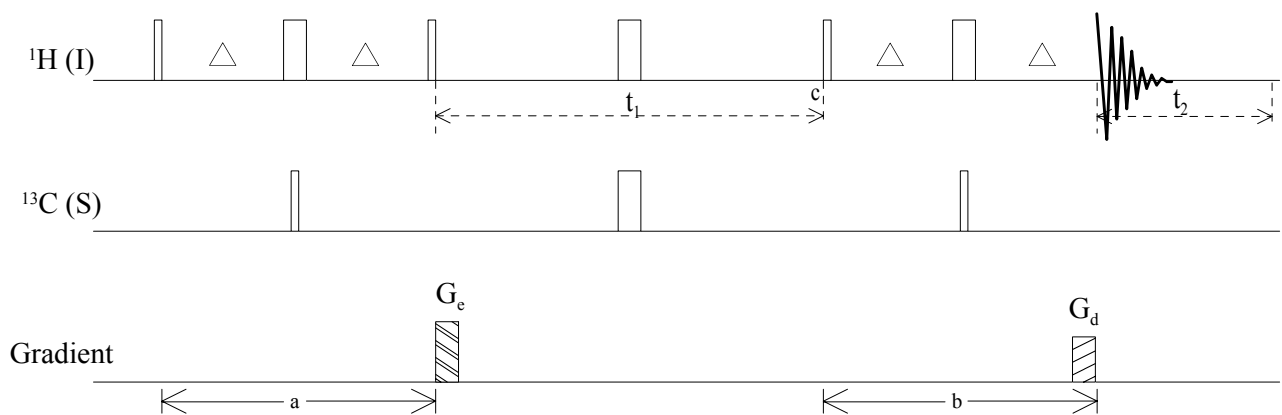
\end{document}